\documentclass[aps,prd,eqsecnum,showpacs,superscriptaddress,
twocolumn,nofootinbib,floatfix,preprintnumbers,altaffilletter]{revtex4}
\usepackage{graphicx}
\usepackage{hyperref}
\usepackage{amssymb}
\usepackage{amsmath}
\usepackage{longtable}
\usepackage{rotating}
\usepackage{color}
\usepackage{fancyhdr}

\def\lesssim{\mathrel{\hbox{\rlap{\hbox{\lower4pt\hbox{$\sim$}}}\hbox{$<$}}}}
\def\gtrsim{\mathrel{\hbox{\rlap{\hbox{\lower4pt\hbox{$\sim$}}}\hbox{$>$}}}}
\def\cos{\rm cos}
\def\sin{\rm sin}

\newcommand{\be}{\begin{equation}}
\newcommand{\ee}{\end{equation}}
\newcommand{\bea}{\begin{eqnarray}}
\newcommand{\eea}{\end{eqnarray}}
\newcommand{\bdm}{\begin{displaymath}}
\newcommand{\edm}{\end{displaymath}}

\begin{document}
\title{An improved, ``phase-relaxed'' ${\mathcal F}$-statistic for gravitational-wave data analysis}
\author{Curt Cutler}
\address{Jet Propulsion Lab, 4800 Oak Grove Dr., Pasadena, CA 91109}
\address{Theoretical Astrophysics, California Institute of
Technology, Pasadena, California 91125}



\date{\today}

\begin{abstract}
Rapidly rotating, slightly non-axisymmetric neutron stars emit nearly periodic gravitational waves (GWs),  quite
possibly at levels detectable by ground-based GW interferometers.
We refer to these sources as ``GW pulsars''. 
For any given sky position and frequency evolution, 
the ${\mathcal F}$-statistic 
is the optimal (frequentist) statistic for the detection of GW pulsars.
However, in an "all-sky" searches for previously unknown 
GW pulsars,
it would be computationally intractable to calculate the (fully coherent) ${\mathcal F}$-statistic
at every point of (suitably fine) grid  covering the parameter space: the number of gridpoints is many orders
of magnitude too large for that.
Therefore, in practice some
non-optimal detection statistic is used for all-sky searches.  
Here we introduce a ``phase-relaxed''
${\mathcal F_{pr}}$-statistic, which we denote  ${\mathcal F_{pr}}$, for incoherently combining the results of fully coherent
searches over short time intervals.  We estimate (very roughly) that for realistic searches, our
${\mathcal F_{pr}}$ is $\sim 10-15\%$ more sensitive than the ``semi-coherent'' ${\mathcal F}$-statistic
that is currently used.  Moreover, as a byproduct of computing ${\mathcal F_{pr}}$, one obtains
a rough determination of the time-evolving phase offset between one's template and the 
true signal imbedded in the detector noise. 
Almost all the ingredients that go into calculating 
${\mathcal F _{pr}}$ are {\it already}
implemented in the LIGO Algorithm Library, so we expect that relatively little additional effort
would be required to develop a search code that uses ${\mathcal F _{pr}}$.

\end{abstract}
\maketitle 

\section{Introduction}
\label{Sec:Int}
 
The ${\mathcal F}$-statistic
is the optimal frequentist statistic for the 
detection of nearly monochromatic gravitational waves (GWs) from a  
neutron star with known (or assumed) sky location and frequency evolution~\cite{jks}.
 The basic idea behind the ${\mathcal F}$-statistic is
simply this: for any given sky location and frequency evolution, 
the set of possible GW signals
forms a four-dimensional (real) vector space~\cite{cf94,jks}. The four basis vectors are the two 
quadratures ($\sin$ and $\cos$) of each of the two polarization bases, $+$ and $\times$.  
Because the set is a vector space (not just a 4-d manifold), it is computationally trivial to maximize the likelihood function over this set;  ${\mathcal F}$ is the maximized log-likelihood. 
For Gaussian noise,  the probability distribution function (pdf) of ${\mathcal F}$ is also particularly simple:
 a  (perhaps non-central) $\chi^2$ distribution with $4$ degrees of freedom.
In their original paper by Jaranowski, Krolak \& Schutz~\cite{jks} (hereinafter referred to 
JKS),  the ${\mathcal F}$-statistic was derived only for the  
case of a single GW detector 
and a single GW pulsar. 
Cutler \& Schutz~\cite{CutlerSchutz2005} showed how the ${\mathcal F}$-statistic can be generalized in a 
straightforward manner 
to the cases of 1) a network of detectors
noise curves, 
and 2) an entire collection of known sources. 

In practice, searches for nearly-monochromatic GWs are separated into a few
different types, depending on how much is know about the source.  The different
types of searches can have vastly different computational requirements.  While the search
for a  GW counterpart to a known radio pulsar is trivial in terms of computational burden,
``all-sky" searches for GW pulsars with no known counterpart (and hence unknown frequency
and frequency derivatives) are currently limited by the available computational power.  
That is, we could dig deeper into the existing data sets if we possessed either larger computational 
resources or more efficient algorithms.   In this paper we demonstrate a way
of significantly improving on the existing algorithms.

Currently, the most sensitive all-sky searches are based on the following idea~\cite{EH}.
For a GW pulsar with unknown frequency evolution (i.e., unknown $f$, $\dot f$, etc.),
computational power required for an optimal (i.e, fully coherent, ${\mathcal F}$-based) 
search grows as a high-power of the total observation time, $T$.  Therefore in practice one divides $T$ into some number N (typically of order $10^2)$ short intervals of duration
$\Delta T = T/N$,  performs an optimal, coherent search one each short interval, and then 
``adds up" the power from the all subintervals.   More specifically, the current method is to calculate a ``semi-coherent''
detection statistic ${\mathcal F}_{sc}$, defined as the sum of the ${\mathcal F}$ values from each of the short intervals:
\be\label{fsc}
{\mathcal F_{sc}}  \equiv \sum_{i=1}^{N} {\mathcal F}_i \, .
\ee
Because calcuting each ${\mathcal F}_i$ involves a maximization over $4$ free parameters, the pdf for ${2\mathcal F_{sc}}$   is a $\chi^2$ distribution with $4N$ d.o.f. 
But this is far more parameters than are actually needed to describe the
physical system!   Consider the very first interval.  The imbedded GW signal is described by $4$ parameters: $(h_0,  \iota, \psi, \Phi)$.  But the triplet $(h_0,  \iota, \psi)$ are the same for all $N$ intervals.  All that changes from interval to
interval is the overall phase $\Phi$.  Assuming maximum ignorance of the signal's phase evolution, one therefore
needs $N-1$ additional phases to fully describe the signal.  So the GW signal is fully described by only $N+3$ parameters.
We define ${\mathcal F_{pr}}$  to be the maximized log-likelihood on this $N+3$-dimensional space.
Our main aims in this paper are (i) to demonstrate an efficient algorithm for calculating  ${\mathcal F_{pr}}$ and (ii) to 
and to illustrate its superiority as a detection statistic 
(superior in the sense of improved ROC curves, where ROC stands for 
"Receiver Operating Characteristics").  

We note that our work is rather similar in spirtit to a recent paper by Dergachev~\cite{Dergachev2010}, though we believe that
we have advanced the idea considerably further.
We note, too, a recent paper by Pletsch~\cite{pletsch2011}, which addresses the followng, rather different weakness in the
detection statistic Eq.~(\ref{fsc}). 
To understand the problem,  
let $(t_0, t_1, \cdots, t_N)$ be the boundary points of the $N$ short intervals,
and consider any one such boundary time, say $t_{50}$.  In Eq.~(\ref{fsc}), data sampled ony slightly earlier than $t_{50}$ 
gets combined {\it incoherently} with data sampled only slightly later than $t_{50}$, which clearly exacts a price in sensitivity.
Pletsch overcomes this problem (which stems from the rather arbitrary choice of boundary points) using his ``sliding window" technique~\cite{pletsch2011}.  We suspect that relatively simple refinements of the detection statistic that we develop in this paper
could also capitalize on Pletsch's basic insight, but we leave such refinements to future work.
%
%
%

The plan of this paper is as follows. In Sec.\ \ref{SecII} we briefly 
review the fully coherent ${\mathcal F}$-statistic, partly to establish notation. 
We generally try to align our notation with that 
of JKS, to ease comparison with their work.  
We also review the (currently used) semi-coherent ${\mathcal F}$-statistic
and its properties.

As further motivation for our work, in Sec.\ \ref{warmup} we consider a ``warm-up'' problem that
we can easily treat analytically and that qualitatively has much in common with our actual problem.
In Sec.\ \ref{example}, in order to illustrate both  the use of  ${\mathcal F_{pr}}$, we present numerical results for
one example search.  For the same search, we also investigate the relative power/sensitivty
of   ${\mathcal F_{pr}}$ versus ${\mathcal F_{sc}}$.  
Our conclusions are summarized in Sec.\ \ref{conclusion}.  

This paper represents our "first-cut" analysis of the ${\mathcal F_{pr}}$ statistic.
There remains significant follow-up work to better elucidate the properties of 
${\mathcal F_{pr}}$ and to implement it in realistic, hierarchical searches.
This future work is also summarized in Sec.\ \ref{conclusion}.

\section{Review of Signal Processing for GW pulsars}
\label{SecII}
In this section we review the rudiments of signal processing that we will require, partly to fix notation.  We also review both the coherent and semi-coherent versions of the ${\mathcal F}$-statistic.  For simplicity of exposition, in 
this paper we will restrict the case of a single detector, and assume that 
the detector noise is stationary. 
The extension to the more realistic case of multiple detectors with slowly changing
noise spectra is completely straightforward.

\subsection{Mathematics of signal processing}
We begin by reviewing the basic mathematics of signal processing. 
For more details, we refer the reader to Thorne (1987), Finn \& Chernoff (1993), and/or Cutler \& Flanagan (1994)~\cite{300years,finn2,cf94}.

Assuming that the noise is stationary and Gassuian, the noise spectral density $S_h(f)$ determines a natural inner product $\left( \ldots | \ldots \right)$ on the
vector space of all detector outputs $x(t)$:
\FL
\begin{equation}
\label{product_def}
\left( {\bf x} \, | \, {\bf y} \right) \equiv 2 \,  \int_{-\infty}^{\infty} df
\,\, \frac{\tilde x^*(f) \tilde y(f)}{S_h(f)} \,   , 
\end{equation}
where $\tilde x(f)$ and $\tilde y(f)$ denote the Fourier transforms of $x(t)$ and $y(t)$,  and $S_h(f)$ is the {\it single-sided} spectral density of the noise. 
In terms of this inner product,
the probability distribution function (pdf) for the noise  ${\bf n}(t)$
takes the form
\begin{equation}
\label{noise_distribution}
{\rm pdf}[{\bf n}] \,= \, {\cal N} e^{- \left( {\bf n} | {\bf n}\right) / 2 },
\end{equation}
\noindent where $\cal N$ is a normalization constant.
Using $\big< \cdots \big>$ to denote "expectation value" (over many realizations of the noise),
it follows from Eq.~(\ref{noise_distribution}) that
\be\label{expectation}
\big< \, \left( {\bf x} \, | \, {\bf n} \right) \left( {\bf y} \, | \, {\bf n} \right) \, \big> = \left( {\bf x} \, | \, {\bf y} \right) \, .
\ee

In this paper, we will be concerned with waveforms $h(t)$ that are nearly monochromatic (here meaning that their frequencies $f(t)$ are slowly varying). In this 
case their inner product is equally simple in the time domain.  Taking the measurement time interval to be $0$ to $T$, we have
\FL
\begin{equation}
\label{product_def}
\left( {\bf h_1} \, | \, {\bf h_2} \right) = 2 \,  \int_0^T 
\,\, \frac{{\tilde h_1}(t){\tilde h_2}(t)}{S_h\big(f(t)\big)}\, dt \,   , 
\end{equation}

\subsection{The fully coherent ${\mathcal F}$-statistic}
\label{multiple}
Next we briefly review the use of the coherent F-statistic in GW pulsar searches.  For more details 
we refer the reader to Cutler \& Schutz (2005)~\cite{CutlerSchutz2005}.
Consider a nearly monochromatic GW signal from an individual 
source with known sky location and known frequency evolution $f(t)$. 
The GW signal is then characterized by
four remaining unknowns: an overall amplitude $A$ (equivalent to the combination
$h_0\ {\rm sin}\ \zeta \,  {\rm sin}^2\theta$ in the notation of JKS), two angles
$\iota$ and $\psi$ that characterize the waves' polarization (equivalent 
to determining the direction of the NS's spin axis), and 
an overall phase $\Phi$.  

The GW signal $h(t)$ registered by the detectgor depends nonlinearly
on $\iota, \psi, \Phi$, but, crucially, one can make a simple change of 
variables--to $\big(\lambda^1, \lambda^2, \lambda^3, \lambda^4 \big)$--such
that dependence of $h(t)$ is linear in these new variables:
\be 
h(t) = \sum_{a=1}^4 \lambda^a h_a(t)
\ee
\noindent where the four basis waveforms $h_a^{\alpha}(t)$ are
defined by
\FL
\bea\label{hdef2}
&&h_1(t) = F_+(t) \cos \Phi(t) , \   
h_2(t) = F_{\times}(t) \cos \Phi(t) , \ \ \ \ \ \ \ \ \ \nonumber \\
&&h_3(t) =  F_+(t) \sin \Phi(t) , \
h_4(t) = F_{\times}(t) \sin \Phi(t)  \, .\ \ \ \ \ \ \ \ \ 
\eea
Here $\Phi(t)$ is the waveform phase at the detector:
\be \label{Phase}
\Phi(t) \approx 2\pi \int^{t} f(t') dt'  \, ,
\ee
where $f(t')$ is the measured GW frequency at the detector at
time $t'$.  The measured frequency includes the Doppler effect from the
detector's motion relative to the source, as well as Einstein and Shapiro
delays associated with the Earth's orbit around the Sun. 
When the GW pulsar is in 
a binary, then $f(t')$ also includes the Roemer, Einstein, and
Shapiro delays associated with that binary orbit. ( We emphasize that 
the known-pulsar searches described here do {\it not} require that the
GW pulsar be isolated, but just that there exists an accurate
timing model for the emitted waves.)
The $F_+(t)$ and $  F_{\times}(t)$ terms in 
Eq.~(\ref{hdef2}) are the 
beam-pattern functions describe the detector's response   
to the $+$ and $\times$ polarizations, respectively. 
We note that the exact form of $F_+(t)$ and $F_{\times}(t)$ 
depends on one's convention for decomposing the waveform into ``plus'' and
``cross'' polarizations; a one-parameter family of choices is possible,
corresponding to the freedom to rotate the axes around the line of sight.
JKS follow the conventions of 
Bonazzola \& Gourgoulhon~\cite{Bonazzola}.

%
Next we define the $4\times 4$ matrix $\Gamma_{ab}$ by 
\be
\Gamma_{ab} \equiv \big(\frac{\partial {\bf h}}{\partial \lambda^a}\, | \, \frac{\partial {\bf h}}{\partial \lambda^b}\big) 
= \big({\bf h_a} \, | \, {\bf h_b}\big)  \, .
\ee
Because both the observation time and $1$ day (the timescale on which
the $F_{+,{\times}}^{\alpha}(t)$ vary) are vastly larger than 
the period of the sought-for GWs (typically $10^{-2}-10^{-3}$ s), 
we can 
replace $\cos^2\Phi(t)$,  $\sin^2\Phi(t)$, and $\cos\Phi(t) \sin \Phi(t)$ 
by their time-averages: $\cos^2\Phi(t),\sin^2\Phi(t) \rightarrow \frac{1}{2}$, while $\cos\Phi(t) \sin \Phi(t) \rightarrow 0$. Then we have
\bea
\Gamma_{11} &\approx& 
\int{F_+(t)\ F_+(t) S_h^{-1}\big(f(t)\big)\, dt} \nonumber \\
\Gamma_{12} &\approx & 
\int{F_+(t)\ F_{\times}(t) S_h^{-1}\big(f(t)\big) \, dt} \nonumber \\
\Gamma_{22} &\approx&  
\int{F_{\times}(t)F_{\times}(t) S_h^{-1}\big(f(t)\big)\, dt} \, ;
\eea
\noindent additionally, $\Gamma_{33} \approx \Gamma_{11}$, 
 $\Gamma_{34} \approx \Gamma_{12}$, $\Gamma_{44} \approx \Gamma_{22}$, and
 $\Gamma_{13} \approx \Gamma_{14} \approx \Gamma_{23} \approx \Gamma_{24} \approx 0$.

The best-fit values of $\lambda^a$ satisfy 
\be
{\frac{\partial }{\partial \lambda^a } } ({\bf x} - \sum_b \lambda^b {\bf h_b}\,  | \, x - \sum_c \lambda^c {\bf h_c}) = 0
\ee
\noindent implying 
\be
\lambda^a = \sum_b \, (\Gamma^{-1})^{ab}({\bf x} \, | \, {\bf{h_b} }) \, .
\ee
\noindent
Then $2{\mathcal F}$, which is defined to be twice the log 
of the maximized likelihood ratio, is just
\bea\label{F1} 
2{\mathcal F} 
&=& ({\bf x} | {\bf x} ) - ({\bf x} - \sum_b \lambda^b {\bf h_b} \ \, | \, \ {\bf x} - \sum_c \lambda^c {\bf h_c}) \nonumber \\
&=& \sum_{a,d} \, (\Gamma^{-1})^{ad}({\bf x}|{\bf h_a}) ({\bf x}|{\bf h_d}) \, .\label{F2}
\eea
\noindent   Using $2{\mathcal F}$ as one's detection statistic satisfies the
Neyman-Pearson criterion for an optimum test: it minimizes the 
false dismissal (FD) probability for any given false alarm (FA) probability. 

Writing ${\bf x} = {\bf n} + {\bf h}$, and plugging into Eq.~(\ref{F2}), we
find 
\be 
\big< 2{\mathcal F} \big> = 4 + ({\bf h}\  |\ {\bf h}) \, ,
\ee
where we have used Eq.~(\ref{expectation}) and the fact that \linebreak
$ \big<({\bf h}\  |\ {\bf n})\big> \ = \  0 $.  More generally, it is easy to show
that $y \equiv 2{\mathcal F}$ follows a $\chi^2$ distribution with
$4$ degrees of freedom (d.o.f) and non-centrality parameter $\rho^2 \equiv 
({\bf h} | {\bf h})$:
\be\label{py}
P(y) = \chi^2(y |4; \rho^2) \, .
\ee

As pointed out by JKS, if we use the following complexified
variables, 
\be\label{FaFb}
2 F_a \equiv ({\bf x} | {\bf h_1} - i {\bf h_3}) \  , \ \ 
2 F_b \equiv ({\bf x} | {\bf h_2} - i {\bf h_4}) \ ,
\ee
then the expression (\ref{F2}) for $2{\mathcal F}$
can be re-written in a particularly simple form:
Eq.~(\ref{F2}) becomes
\be\label{complex}
2{\mathcal F} = 
\frac{8}{D}\big[\negthinspace B|F_a|^2 + A|F_b|^2 - 
2C \Re (F_a F^*_b)\big] \, .
\ee
\noindent where
\be 
A \equiv \,  ({\bf h_1}|{\bf h_1}) \,  , \ \ \ B \equiv   ({\bf h_2}|{\bf h_2})
\, , \ \ \ C \equiv  ({\bf h_1}|{\bf h_2})  \, ,
\ee
\noindent and $D \equiv AB - C^2$.
\noindent (Note that the A,B,C terms defined here are, in the single-detector case, 
larger than the A,B,C terms in JKS by a factor of the observation time
$T$.)

\subsection{The ``semi-coherent'' ${\mathcal F}$-statistic}\label{semi}
As mentioned above, the current method of incoherently combining the coherent results from successive intervals
is just to sum of the  ${\mathcal F}$-statistics from all the intervals:

\be\label{sum}
2{\mathcal F_{sc}} 
\equiv  \sum_{i=1}^{N} 2{\mathcal F_i} \, .
\ee 
It also easy to show that $y \equiv 2{\mathcal F_{sc}}$ follows a $\chi^2$ distribution with $4N$ degrees of freedom:
\be\label{pym}
P(y) = \chi^2(y |4N; \rho^2) \, .
\ee
where the non-centrality parameter $\rho^2 = \sum_{i=1}^{N} \rho^2_i$.

In the cases of interest to us, $4N$ will generally be large, and then the $\chi^2$ distribution 
with $4N$ d.o.f. can often be approximated as a Gaussian.
Let $y \equiv 2{\mathcal F}$, and let $\rho_{tot}^2 \equiv \sum_{i=1}^{N}\rho_{i}^2 $.
Then 
\be
P(y) = \chi^2(y|4N;\rho_{tot}^2) \approx (8\pi\, N)^{-1/2} e^{-(y - <y>)^2/(8N)}
\ee
\noindent where $<y> = 4N + \rho_{tot}^2 $. For example, using this approximation (and the fact that for a Gaussian $P(y)$,  events $2.326 \sigma$ above the mean
occur $1\%$ of the time), we see that 
the threshhold value $y_{th}$ that yields a $1\%$ FA probability is
%
\be\label{t1}
y_{th} \approx  4N + 2.326 \sqrt{8N}  \ \  {\rm (large \,  N)}  \, .
\ee


\section{The ``phase-relaxed'' ${\mathcal F}$-statistic}
\label{Fpr}
We are now ready to define ${\mathcal F_{pr}}$.  Basically, 
${\mathcal F_{pr}}$ coincides with the full matched-filtering $SNR^2$, under the
assumption that the manifold of waveforms is $N+3$-dimensional (i.e., 4 parameters for
the first segment, and $N-1$ for the relative phase offsets of the remaining segments).
What makes ${\mathcal F_{pr}}$ useful in practice is that we have also found a simple and efficient method for calculating it.

\subsection{Motivation and definition}

We begin by defining complex basis functions $H_+$ and $H_{\times}$ by
\be\label{Hdef}
H_+ \equiv h_1 - i h_3 \ \ \ \ , \ \ \  H_{\times} \equiv h_2 - i h_4 \, . 
\ee

This complex representation is especially convenient for our purposes because
$H_+$ and $H_{\times}$ both transform very simply under an overall phase shift in $\Phi(t)$:   under $\Phi(t) \rightarrow \Phi(t) + \delta$,  $H_+$ and $H_{\times}$ 
transform as 
$H_+(t) \rightarrow e^{- i \delta} H_+(t)$ and $H_{\times}(t) \rightarrow e^{- i \delta} H_{\times}(t)$.  (Note that the minus sign in the exponent 
in the term $e^{- i \delta}$ stems from the minus signs in the definitions of
of $H_+$ and $H_{\times}$ in Eq.(~\ref{Hdef}).
For these complexified signals, our usual inner product becomes a Hermitian one;
for nearly monochromatic signals near frequency $f$, this Hermitian
inner product is given simply by

\FL
\begin{equation}
\label{product_def}
\big( {\bf x} \, | \, {\bf y} \big)  = \frac{2}{S_h(f)} \int{x^*(t) y(t) dt} \ \ \, .
\end{equation}
\noindent Clearly $ \big( {\bf x} \, | \, {\bf y} \big)  = \big( {\bf y} \, | \, {\bf x} \big)^*$.

Next we define $\Gamma_{\alpha \beta}$ by 
\be\label{gamdef}
\Gamma_{\alpha \beta} \equiv \big(H_{\alpha}\, | \, H_{\beta} \big) = 
\Gamma^*_{\beta \alpha} \, ,
\ee
\noindent where $\alpha$ and $\beta$ run over $+,\times$.
It follows immediately that for GW data $x(t)$, (twice the) ${\mathcal F}$-statistic is given by
\be\label{cplx-fstat}
2 {\mathcal F} = \big(\Gamma^{-1}\big)^{\alpha \beta} \big( H_{\alpha} \, | \, {\bf x} \big) \big( {\bf x} \, | \, H_{\beta} \big) \, .
\ee

Now imagine breaking up the full integration time $T$ into N intervals
of duration $\Delta T_i$, for $i = 1, 2, \cdots , N$.  (We expect that 
in practice the $\Delta T_i$ will generally be of approximately the same length,
but this is not required.)  Next define $ x^i(t)$ to be
the restriction of $x(t)$ to the $i^{th}$ interval; i.e,
$x^i(t) = x(t)$ for $t$ in the $i^{th}$ interval, and $x^i(t) = 0$ for $t$ outside the $i^{th}$ interval.  Then clearly we have 
\be\label{cplx-fstat2}                                                           
2 {\mathcal F} = \big(\Gamma^{-1}\big)^{\alpha \beta} \bigg(\sum_i \big( H_{\alpha} \, | \, {\
\bf x^i} \big) \big( \sum_j\big( {\bf x^j} \, | \, H_{\beta}\big) \bigg)\, .               
\ee                                                                    

To motivate our definition of ${\mathcal F}_{pr}$, recall
that if we had practically limitless computer power at our disposal, then the most sensitive search would be a coherent
matched-filter search over a fine grid covering the entire GW-pulsar parameter space.  However for ``blind'' GW pulsar searches (i.e., searches for GW pulsars whose sky location and/or time-changing frequency are unknown), maintaining phase coherence between the template signal and true imbedded signal, over timescales of months to years, would require an extremely  fine grid on parameter space, and (one easily shows) many  of orders of magnitude more computing power than is realistic~\cite{BCCS}

The basic idea behind ``semi-coherent'' searches is to employ a detection statistic
that is less sensitive to phase decoherence across the whole observation time,    
which allows one to use a much coarser grid on
parameter space.  In effect, one sacrifices some sensitivity in the interest of computational practicality. 
For our phase-relaxed ${\mathcal F}$-statistic, the idea is that the 
search-template signal should 
remain approximately in phase with the true, imbedded signal in each interval $\Delta T_i$--up to some constant phase ``offset
$\delta_i$--but that the $\delta_i$ should be allowed to vary from interval to interval.
That is, we replace 
\bea\label{cplx-fstat3}                                                                                                   
& \big(\Gamma^{-1}\big)^{\alpha \beta} \bigg(\sum_i \big( H_{\alpha} \, | \, {\
\bf x^i} \big) \bigg) \bigg( \sum_j \big({\bf x^j} \, | \, H_{\beta} \big) \bigg)\, \rightarrow  \nonumber \\
& \big(\Gamma^{-1}\big)^{\alpha \beta} \bigg(\sum_i \big(H_{\alpha} \, | \, {\
\bf x^i}\big)  e^{i \delta_i}\bigg) \bigg( \sum_j \big({\bf x^j} \, | \, H_{\beta}\big)  e^{-i \delta_j} \bigg)\, .     
\eea
Finally, we define (twice) ${\mathcal F_{pr}}$ to be the rhs of (\ref{cplx-fstat3}), maximized over all phase-offsets
$\delta_i$:
\be\label{cplx-fstat4}                                                                                                     
2 {\mathcal F}_{pr}  = \max_{\delta_1, \cdots, \delta_n}\big\{  \big(\Gamma^{-1}\big)^{\alpha \beta} \bigg(\sum_i \big( H_{\alpha} \, | \, {\                                                
\bf x^i} \big) e^{i \delta_i}\bigg) \bigg( \sum_j\big( {\bf x^j} \, | \, H_{\beta} \big) e^{-i \delta_j} \bigg)\big\} \, .     \
\ee                                                                                                                  

While there are $N$ phase angles $\delta_i$, only $N-1$ of them are actually independent; i.e., it is easy to check that
${\mathcal F_{pr}}$ is invariant under $\delta_i \rightarrow \delta_i + c$, where $c$ is any constant.

\subsection{Maximizing over the phase offsets $\delta_i$}

The whole point of developing alternatives to the fully coherent  ${\mathcal F}$-statistic is
to save on computational cost, so for our phase-relaxed ${\mathcal F}$-statistic to be 
useful, we need a reasonably efficient way of maximizing over the $\delta_i$.  In this section
we demonstrate one efficient method.  We demonstrate only the simplest version of this
method, which we regard as basically an ``existence proof'' that efficient methods do exist.
It should be clear by the end of this section that there are many variations on our basic method by
which one might attempt to improve its efficiency, but we defer such improvements to later work.

Our method is as follows.   We can simplify the appearance of the equations by defining
\be \label{kij-def}
K^{ji} \equiv \big(\Gamma^{-1}\big)^{\alpha \beta} \big( {\bf x^j} \, | \,
H_{\beta}\big) \, \big(H_{\alpha} \, | \, {\bf x^i} \big) 
\ee
and defining $v$ to be the following $N$-dimensional vector formed out of the phase offsets: 
\be\label{vdef}
{\bf v} \equiv (e^{i \delta_1},  e^{i \delta_2},  \cdots e^{i \delta_n} ) \, .
\ee
\noindent Note that
$K^{ji}$ is Hermitian (i.e., $K^{ij} = K^{ji\, *}$),
and that we can now re-write  Eq.~(\ref{cplx-fstat3}) as
\be\label{cplx-fstat4}                                                                                                     
2 {\mathcal F}_{pr}  = \max_{\delta_1, \cdots, \delta_n}  {v_j}^* K^{ji} v_i \, .
\ee
Of course, $K^{ji}$ is completely determined by the two complex templates $H_{\alpha} $ and their inner products with
the data, while our goal is to find the $v_i$ that maximize $v_j^* K^{ji} v_i$, subject to the N constraints that  $v_i {v_i}^* = 1 \forall i$.  (We emphasize that $i$ is {\it not} summed over in these constraints.)
Put another way: ${\bf v}$ lies on the unit N-torus (i.e, the unit circle cross itself N times.)  Naturally,  we employ the method of 
Lagrange multipliers to maximize $v_j^* K^{ji} v_i$ on this constraint surface.  Since there are N constraints, we obtain N
equations with N (real) Langrange multipliers $\lambda_j$:
\be\label{Lag} 
K^{ji} v_i = \lambda_j v_j  \ \ \forall  j   \, .
\ee
\noindent We emphasize that Eq.~(\ref{Lag}) is {\it not} an eigenvalue equation, since in general the 
$N$ values $\lambda_j$ will all be different.  

Next we find it convenient to introduce a projection operator P operating on ${\mathbb C}^N$.
Let ${\bf w} = (c_1 e^{i \delta_1}, c_2 e^{i \delta_2}, \cdots , c_N e^{i \delta_N} )$, where the $c_i$ are all real. Then P is defined by
\be
P {\bf w} =  (e^{i \delta_1}, e^{i \delta_2}, \cdots , e^{i \delta_N} ) \, . 
\ee
\noindent I.e., the operator P takes any vector in ${\mathbb C}^N$ and projects it down onto the unit torus.  (Note that P is {\it not}  a
linear operator, but it is true that $P^2 = P$.)
Then Eq.~(\ref{Lag}) is clearly equivalent to the requirement that 
\be\label{Lag2}
P K {\bf v} = {\bf v}  \, .
\ee

Hence the solution ${\bf v}$ is a fixed point of the operator  $P K$.  In fact, numerical experience shows that it is an attractive fixed point.
That is, let  ${\bf v_0} $ 
be some initial guess, and then operate on it repeatedly with $P K$.  Define
$(P K)^2 \equiv (P K) (PK)$,  $(P K)^3 \equiv (P K) (PK) (P K)$, etc.   Then for ${\bf v_0}$ sufficiently
close to the true solution ${\bf v_0}$, we find that
\be
(P K)^m {\bf v_0} \rightarrow {\bf v}  \, 
\ee
\noindent as $m$ increases.  In practice, we find that the convergence is quite rapid, and
that the initial guess ${\bf v_0}$ need not be particulary close to the solution ${\bf v_0}$.
In numerical experiments (in many thousands of cases, and covering a large range of $N$) we found  that the following initial guess always led to converge of the iterated sequence. 
For each segment $\Delta T_i$, it trivial to calculate the fully coherent ${\mathcal F}_i$
and the corresponding best-fit parameters for that segment alone: $(A_i , \iota_i , \psi_i 
\Phi_i)$.  Then we take as our initial guess
\be
{\bf v_0} = (e^{i \Phi_1},  e^{i \Phi_2},  \cdots e^{i \Phi_n} ) \, ;
\ee
\noindent i.e., the initial guess for the phase offset in each segment is the best-fit
offset for that segment by itself.

\section{Analytic results for a related, warm-up problem}\label{warmup}
It is common sense that when one goes to solve some problem numerically, it is useful to
have analytical results with which to compare it--ideally for a special case of the true problem, or, failing that,
for some qualitatively similar problem.  In this section we derive analytic results for the following case:
Consider a vector space of waveforms that is completely described by $2$ parameters per interval--so $2N$ parameters in all, where $N$ is large--and consider two different searches: one search that maximizes the fit over those $2N$ parameters, and another, less efficient search, that begins with a $4N$-dimensional vector space (in which the true, $2N$-dim vector space lies), whose detection statistic is the maximized log-likelihood on the $4N$-dimensional space.
That is, our two detection statistics are the $2N$- and $4N$-dimensional ${\mathcal{F}}$-statistics, which in this section
we will denote ${\mathcal{F}_{2N}}$ and ${\mathcal{F}_{4N}}$.   

For each search, there is a threshold value $\rho_{th}$ such that the signal is detectable with $FA = 0.01$ and 
$FD = 0.5$.  We can solve both problems at the same time, by considering the general M-dimensional search.
Then the expectation value of ${\mathcal{F}_{M}}$ is $\big<{\mathcal{F}_{M}}\big> = M + \rho^2$ and 
its standard deviation is $\sigma_M = (2M + 4\rho^2)^{1/2}$.   For large $M$, the $\chi^2$ function approaches
a Gaussian, so we will approximate the pdf of ${\mathcal{F}_{M}}$ as a Gaussian with this mean and standard deviation.
Then the threshold for detection with $FA = 0.01$ is 
\bea
\rho^{th}_M &=& \big<{\mathcal{F}_{M}} \big>+ \sqrt{2} \sigma_M\  {\rm erfc}^{-1}(2\, FA) \\
 &=&   M + 3.29 M^{1/2} \,   ,
\eea
\noindent where in Eq.~(\ref{rhoth}) both ${\mathcal{F}_{M}}$  and $\sigma_M$ are 
to be evaluated at $\rho = 0$. Therefore $\rho^{th}_M = \sqrt{3.29 M^{1/2} } = 1.814 M^{1/4}$, and we have
\be
\rho^{th}_{4N}/\rho^{th}_{2N}   =  2^{1/4} = 1.189 \, . 
\ee
\noindent Therefore using the correct statistic allows one to see sources $19 \%$ farther away.

For comparison with results in the next section, we also plot in 
Fig.~1  the FA  vs. FD curves for the two statistics, for a range of $\rho$ values.  
\begin{figure}\label{d2d4}
\centerline{\includegraphics[width=0.55\textwidth]{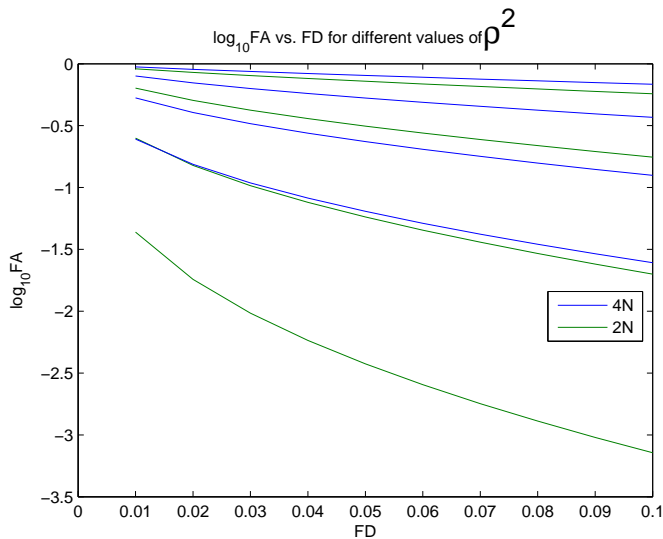}}
\caption{Compares the false alarm (FA) probabilities for the two detection statistics, ${\mathcal{F}_{4N}}$ and ${\mathcal{F}_{2N}}$,
as a function of false dismissal (FD) probability, for several values of the squared signal strength, $\rho^2$. The blue
curves are for ${\mathcal{F}_{4N}}$ and the green for ${\mathcal{F}_{2N}}$.  From upper to lower, the squared signal strengths
are $\rho^2 = 25, 50, 75, 100$.}
\end{figure}

\section{Numerical results for one example search}\label{example}
To illustrate the utility of our ${\mathcal F_{pr}}$ statistic, in this section we present results for a simple, one-parameter family of examples (where the
varied  parameter is the strength of the embedded GW signal),  and we compare the
the effectiveness of ${\mathcal F_{pr}}$ and  ${\mathcal F_{sc}}$. 

We fix the number of intervals 
at $N = 100$, and evaluate search effectiveness for signals with a range of total $\rho^2 \equiv \sum_{i=1}^N \rho^2_i$.
We will eventually consider a range of $\rho$, but for now imagine  $\rho$ as fixed.   For simplicity, in this example 
we will consider a 
case where the $\rho_i$ are the same for all $i$, so  $\rho^2_i = \rho^2/N$,
and where the $\Gamma^i_{ab}$ matrices are also the same for all $i$:  $\Gamma^i_{++} = 3$,
$\Gamma^i_{+\times} = 1=\Gamma^i_{\times +}$ and $\Gamma^i_{\times \times} = 1$ for all $i$.

We decompose the measured signal $x^i$ into waveform plus noise,
\be\label{hn}
x^i = h^i + n^i  \, .
\ee
\noindent
For each $i$, filtering the data with 
$H_+$ and $H_x$ produces two complex numbers:  
$ c^i_+ \equiv \big({\bf x^i} \, | \, H_+ \big)$ and 
$c^i_{\times} \equiv \big( {\bf x^i} \, | \, H_{\times} \big)$.
Clearly, the measured signals $c^i_\alpha$ can be decomposed as
\bea\label{x}
 c^i_{\alpha} &=&\big( {\bf h^i} \, | \, H_{\alpha} \big) + \big({\bf n^i} \, | \, H_{\alpha} \big) \\
 &\equiv& g^i_\alpha  + m^i_\alpha \, .
\eea
\noindent 
We simulate the noise piece $m^i_\alpha$ by taking random draws of (pairs of) complex numbers from a Gaussian distribution with covariance matrix
\bea\label{cov}
\big< {m^{i}_\alpha}^* m^i_\beta \big> &\equiv&  \big< (H_{\alpha} \, | {\bf n^i} \, ) \ \  < ({\bf n^i} \, | \, H_{\beta}) \big> \nonumber \\
&=&  (H_{\alpha} \  \, | \, H_{\beta}) = \Gamma_{\alpha \beta}
\eea
Again for simplicity, we will consider a case where
the $g_i^\alpha$ are the same for each $i$, modulo a random, complex phase factor.
Our particular (and rather arbitrary) choice is  
\be\label{g}
[\ g_i^+ \, , \,  g_i^{\times}]  = (2N)^{-1/2} \rho \ [\ 2 + \sqrt{6} \, , \,  \sqrt{6}  ] \,  e^{i \varphi^i} \, ,
\ee
where the $\varphi^i$ are random phases drawn uniformly from  $[0, 2\pi)$.
One easily checks that $\sum_{i=1}^N (h^i\,  | \, h^i )= \rho^2$.
The inclusion of the $e^{i \varphi^i}$ terms reflects our goal of modeling a case where 
frequency evolutions of the tempate and the true signal are  so mismatched that their relative phases jump significantly and randomly  from one interval to the next.  Choosing the $\varphi^i$ randomly corresponds to the "worst-case scenario", where the true-versus-template phase offsets show no pattern.  In practice, we expect that
the situation will often be much more favorable for searches: i.e., the phase offsets might very often be well fit by some low-order polynomial in time.  
In a later paper we plan to investigate the extent to which the time-evolution of the offsets can be fit by a few parameters, and how that information can be exploited to 
speed up other parts of the search.  
%
%

Given one simulated data set $c^i_{\alpha}$ (200 complex numbers), we compute $2{\mathcal F_{pr}}$.
We repeat for 10000 data sets to determine the distribution of $2{\mathcal F_{pr}}$, and calculate its
mean $\big<2{\mathcal F_{pr}}(\rho)\big>$, standard deviation $\sigma_{pr}(\rho)$, skewness, and kurtosis.  In practice, we find that the skewness and kurtosis are
relatively small 
(as might be expected, since our N is large), so for the rest of this section we will approximate $p( 2{\mathcal F_{pr}}; \rho^2)$ as simply a Gaussian with our measured
mean and standard deviation.  Given these distributions it is completely straightforward to 
determine the false alarm probability $FA$ for any threshold value $2{\mathcal F^{th}_{pr}}$, and
to calculate the false dismissal probability  $FD$ for any pair of $2{\mathcal F^{th}_{pr}}$ and $\rho^2$.

We expect that, in practical searches, ${\mathcal F^{th}_{pr}}$ will find its main use in hierarchical search algorithms, 
in which a very coarse search at relatively low threshold identifies candidates for further examination,
and these are winnowed down in successive stages~\cite{bc,cgk}.  In this context, one generally wants a fairly small
FD rate ($< 1\%$, say), so as not to lose any events,
and strongly prefers a very low FA probability, to 
reduce the computational cost of follow-ups.  With this application in mind, in Figs. 2 and 3. we plot FA as a
function of FD, for several values of $\rho^2$.  
%

\begin{figure}\label{fafd2}
\centerline{\includegraphics[width=0.55\textwidth]{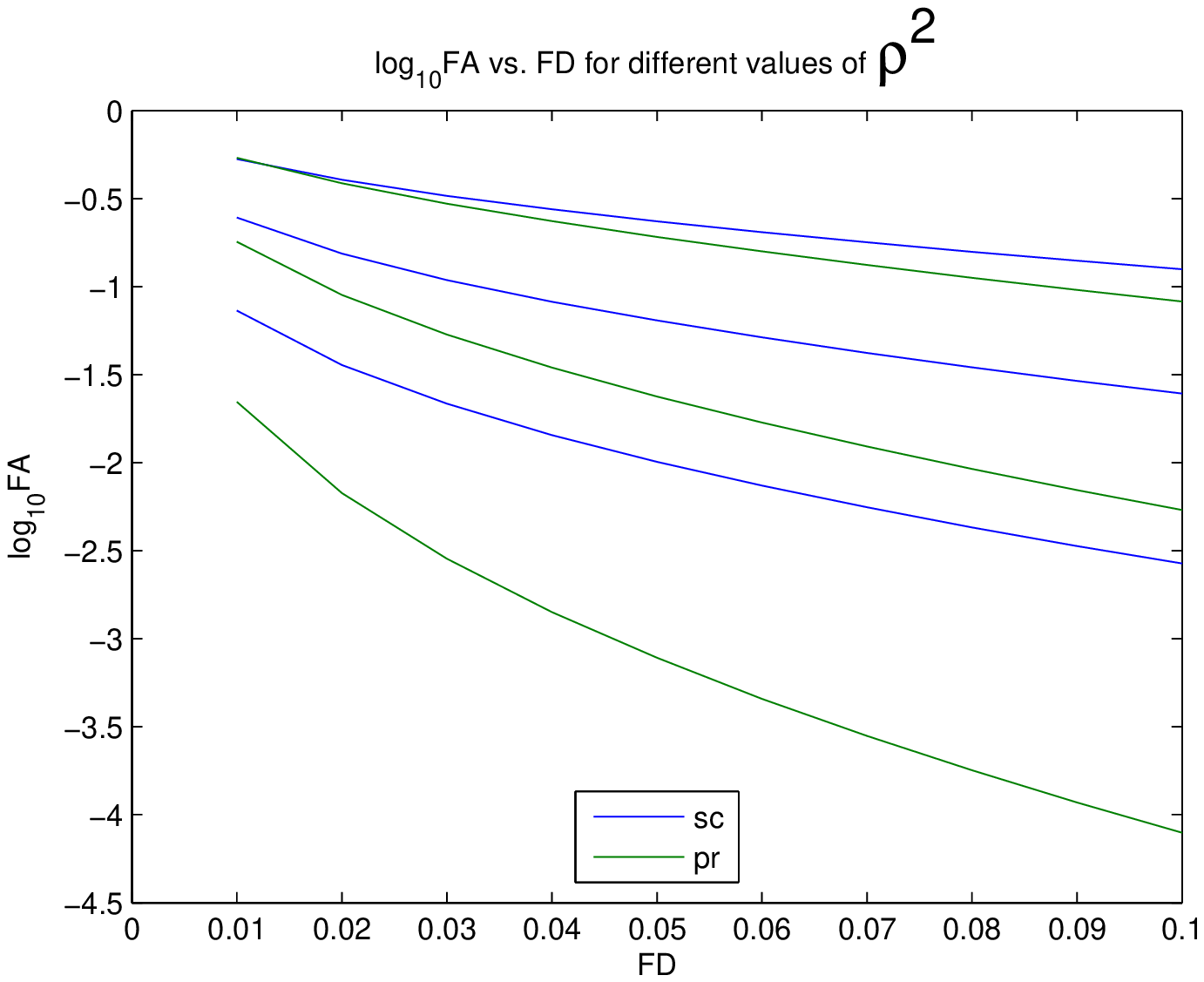}}
\caption{Compares the false alarm (FA) probabilites for the two detection statistics, ${\mathcal{F}_{sc}}$ and ${\mathcal{F}_{pr}}$,
as a function of false dismissal (FD) probability, for several values of the (square of) the signal strength, $\rho^2$. The blue
curves are for ${\mathcal{F}_{sc}}$ and the green for ${\mathcal{F}_{pr}}$.  From upper to lower, the signal strengths
are $\rho^2 = 75, 100, 125$.}
\end{figure}

\begin{figure}\label{fafd3}
\centerline{\includegraphics[width=0.55\textwidth]{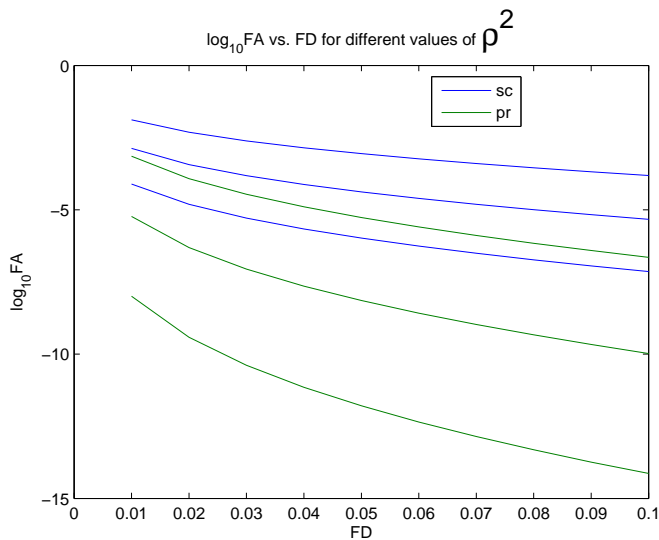}}
\caption{Same as in Fig.~2, except that, from the upper to lower curves, the signal 
strengths are $\rho^2 = 150, 175, 200$.}
\end{figure}

How much does employing ${\mathcal{F}_{pr}}$ increase the sensitity of a blind GW pulsar search?   Obtaining a useful and accurate answer to this question is much more complicated than it might initially seem, since the most sensitive known 
search algorithms for GW pulsars are 
hierarchical searches~\cite{bc,cgk}.   These searches involve  involve several stages,  
with successive  stages "ruling out" an ever-increasing fraction of the parameter space.
We imagine that the most sensitive search--at some fixed, realistic computational cost--might use ${\mathcal{F}_{pr}}$ for only some of its stages.  
And while calculating ${\mathcal{F}_{pr}}$ clearly requires more floating point operations
than calculating ${\mathcal{F}_{sc}}$, it is premature to compare these costs in a detailed way,
since (i) there has as yet been no attempt to speed up our iterative relaxation scheme, and
(ii) the extra information that comes with the phase offsets $\delta_i$ can presumably be
used to speed up other parts of the search.  Despite these difficulties we can obtain a rough
estimate of the sensitivity improvement from using $\rho^2_{pr}$,  as follows.  We have seen that given some detection statistic ${\mathcal  S}$, one naturally obtains a map from 
 the total $\rho^2$ to curves in the $FA-FD$ plane.
 Let ${\mathcal C}({\mathcal  S};\rho^2)$ denote that
curve.  Then we can look for pairs $\rho^2_{pr}$ and $\rho^2_{sc}$
 such that ${\mathcal C}({\mathcal  F_{pr}};\rho_{pr}^2)$ lies close to  ${\mathcal C}({\mathcal  F_{sc}};\rho_{sc}^2)$.  Three such pairs are shown in Fig.~4.    We see that, in the
most relevant portion of $FD-FA$ plane,  ${\mathcal C}({\mathcal  F_{pr}}; 140)$ lies close to  ${\mathcal C}({\mathcal  F_{sc}}; 170)$,   ${\mathcal C}({\mathcal  F_{pr}}; 160)$ lies close to  ${\mathcal C}({\mathcal  F_{sc}}; 200)$, and  ${\mathcal C}({\mathcal  F_{pr}}; 190)$ lies close to  ${\mathcal C}({\mathcal  F_{sc}}; 250)$.  Thus, based on this example, we might estimate that using ${\mathcal  F_{pr}}$ rather than ${\mathcal  F_{sc}}$ affords an increase in sensitivity of $\sim 20-30\%$ in $\rho^2$, or $\sim 10-15\%$ in $\rho$. 
Clearly, to obtain a more reliable estimate we should perform a Monte Carlo simulation
(based on random locations of the source on the sky, and random orientations of
the GW pulsar's spin axis).  We plan to do this in follow-up work.

Comparing Fig.~1 to Fig.~2, we see that the sensitivity gain  
from replacing the detection statistic  ${\mathcal  F_{sc }}$ with ${\mathcal  F_{pr}}$
is qualitatively similar to the gain from ${\mathcal  F_{4N }} \rightarrow {\mathcal  F_{2N}}$,
but that the latter gain is greater (at least for our one numerical example, and for total SNR $\sim 10$).  That may seem surprising, since in the former case we are eliminating $3N-3$
redundant parameters, while in the latter case we are eliminating only $2N$ redundant parameters.  We conjecture that the main reason that the replacement  
${\mathcal  F_{sc }} \rightarrow {\mathcal  F_{pr}}$ "buys us less" in sensitivity is the following.  
In calculating $ {\mathcal  F_{2N}}$, the noise contributions from different intervals $i$ still get
combined incoherently.  However in ${\mathcal  F_{pr}}$, the maximization over the 
phase offsets $\delta_i$ allows the noise contributions to combine coherently.  Indeed
for $\rho_i \alt 1$, the offsets $\delta _i$ are determined more by the noise than by the
imbedded waveform.  However we leave a thorough investigation of  this effect to future work.

\begin{figure}\label{fafd3}
\centerline{\includegraphics[width=0.55\textwidth]{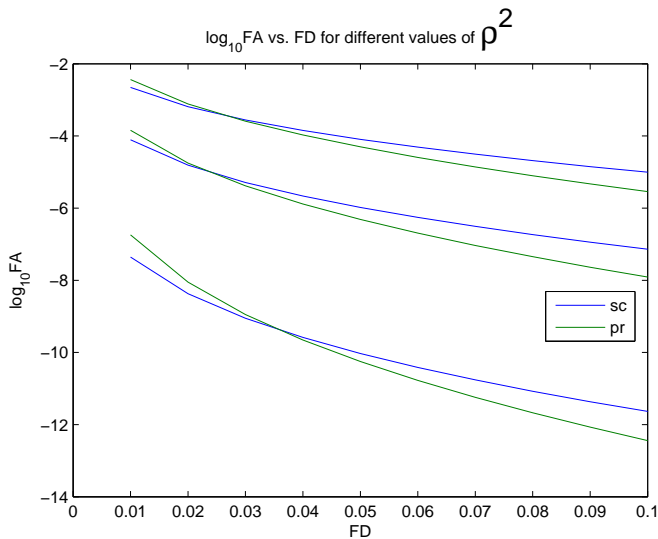}}
\caption{Similar to Figs. 2 and 3, except that here the sensitivity of the ${\mathcal{F}_{sc}}$
statistic is compared to that of ${\mathcal{F}_{pr}}$ at {\it lower} $\rho^2$.
 From upper to lower, the signal strengths for ${\mathcal{F}_{sc}}$ 
are for $\rho^2 = 140, 160, 190$.}
\end{figure}

\section{Summary and Future Work}
\label{conclusion}

In this paper we defined a ``phase-relaxed''  ${\mathcal{F}}$-statistic, denoted ${\mathcal{F}_{pr}}$,
to be used in cases where the total observation time is sufficiently long that straightforward calculation of the fully coherent ${\mathcal{F}}$-statistic over the relevant parameter space is computationally intractable.  The calculation of ${\mathcal{F}_{pr}}$ takes as input the results from
coherent searches over $N$ shorter time intervals. Our ${\mathcal{F}_{pr}}$ coincides with the fully coherent ${\mathcal{F}}$-statistic under the approximation that the phase offsets
between template and imbedded signal are treated as an additional $N-1$ independent parameters.   We also demonstrated one efficient, iterative method for calculating ${\mathcal{F}_{pr}}$.  We regard our iterative method as an "existence proof" for efficient algorithms.  In future work we intend to explore variations on our basic method that
we suspect would lead to substantial improvements in computational cost.  We illustrated the
use of ${\mathcal{F}_{pr}}$ in one simple family of examples, in which the sensitivity improvement (compared to  ${\mathcal{F}_{sc}}$) was shown to be $\sim 10-15\% $.

Our example was based on the "worst-case" assumption that where the phase offsets $\delta_i$ are completely random.  In follow-on work we intend to examine the more realistic case where
the $\delta_i$ are a (reasonably low-order) polynomial in time, and we plan to calculate the increased sensitivity based on a very large number of cases, in Monte Carlo fashion.

Other follow-on projects that we intend to work on include i) the development of new
vetoes~\cite{Itoh} for instrumental artifacts, since, e.g., for true GW pulsars the parameters $(h_0, \iota, \psi)$ calculated from the first half of the
observation should be consistent with those calculated from the second half; ii) the use of the
$\delta_i$ to quickly converge on improved estimates for the GW pulsar's frequency and
spindown parameters, and iii) the optimal use of ${\mathcal{F}_{pr}}$ in multi-stage,
hierarchical searches.

Finally, while our primary interest in this paper has been the application of the ${\mathcal{F}_{pr}}$-statistic to GW pulsars,  it has not escaped our notice that the same idea and formalism can be applied, with only trivial modifications, to searches for (quasi-circular, non-precessing) inspiraling binaries.  With binaries, we expect ${\mathcal{F}_{pr}}$ to be most useful in those cases where the observed GW signal has a very large number of cycles; e.g., searches for
neutron-star binaries by proposed GW detectors that have reasonable sensitivity at $\sim 1$ Hz, such as the 
Einstein Telescope, Decigo, or the Big Bang Observer. 
%
\vskip 0.2in
\begin{acknowledgments}
\vskip -0.2in
This work was carried out at the Jet Propulsion Laboratory, California Institute of Technology, under
contract to the National Aeronautics and Space Administration. We gratefully acknowledges support from
NSF Grant PHY-0601459. 
We also thank Michele Vallisneri, Holger Pletsch, Reinhard Prix, Badri Krishnan and Bruce Allen 
for helpful discussions.   Copyright 2011.  All rights reserved.
\end{acknowledgments}

\end{document}